\documentclass[prd,twocolumn,floatfix,nofootinbib,preprintnumbers,letterpaper]{revtex4}

\usepackage{epsfig} 
\usepackage{amsmath}

\newcommand{\beq}{\begin{equation}}
\newcommand{\eeq}{\end{equation}}
\newcommand{\beqa}{\begin{eqnarray}}
\newcommand{\eeqa}{\end{eqnarray}}

\newcommand{\lcdm}{$\Lambda$CDM}

\begin{document} 

\title{Aetherizing Lambda: Barotropic Fluids as Dark Energy}
\author{Eric V.\ Linder}
\affiliation{Berkeley Lab \& University of California, Berkeley, CA 94720}  
\author{Robert J.\ Scherrer}
\affiliation{Department of Physics and Astronomy, Vanderbilt University,
Nashville, TN  ~~37235}

\date{\today}

\begin{abstract} 
We examine the class of barotropic fluid models of dark energy, in which 
the pressure is an explicit function of the density, $p = f(\rho)$.  
Through general physical considerations we constrain the asymptotic past 
and future behaviors and show that this class is equivalent to the sum 
of a cosmological constant and a decelerating perfect fluid, or ``aether'', 
with $w_{AE}\ge0$.  Barotropic models give substantially disjoint 
predictions from 
quintessence, except in the limit of $\Lambda$CDM.  
They are also interesting in that they simultaneously can ameliorate the 
coincidence problem and yet ``predict'' a value of $w\approx-1$. 
\end{abstract} 


\maketitle

\section{Introduction \label{sec:intro}}

Observational evidence strongly points to an accelerated expansion of the
universe \cite{Komatsu,Kowalski}, but the physical origin of this 
acceleration is unknown.  Since general relativity relates spacetime 
curvature (and hence acceleration) to energy, 
it is natural to hypothesize that either this relationship must be
modified (as in extended gravity models \cite{dgp,fR}), or that there is 
some additional
source of energy density driving the expansion.  Many models in the latter
category have been proposed, including scalar field models (``quintessence")
\cite{ratra,wett88,turner,caldwelletal,liddle,Stein1},
scalar field models with modified kinetic terms (``k-essence")
\cite{Arm1,Garriga,Chiba1,Arm2,Arm3,Chiba2,Chimento1,Chimento2,Scherrerk,dePL}, 
and even more exotic possibilities (see, for example, the review
of \cite{Copeland}). 

Here we examine in more detail a class
of models called barotropic fluids, in which the dark energy pressure $p_{DE}$
is given as an explicit function of the density $\rho_{DE}$.  This can be 
viewed as a very simple prescription, in contrast to other models (such as 
the ones above) where the relation is implicit, written in terms of 
intermediate variables.  Although some barotropic models are 
well studied, we find here a number of new, general properties that makes this 
class of interest in dark energy physics.   Specific models investigated 
previously include the Chaplygin gas \cite{Kamenshchik,Bilic} and 
the generalized Chaplygin gas \cite{Kamenshchik,Bento}, the linear 
equation of state 
\cite{linear97,linear98,linear1,linear2} and the affine equation of 
state \cite{AB,Quercellini}  
(note these are actually the same model), the quadratic 
equation of state \cite{odin04,AB}, and the van der Waals equation of state
\cite{VDW1,VDW2}.  Such models have been considered either as unified models
for dark matter and dark energy together, or as models for the dark
energy alone.  We confine our attention to the latter case. 

One of the key properties of barotropic fluids is that the sound speed, 
$c_s^2 = dp_{DE}/d\rho_{DE}$, does not have to equal the speed of light 
as in quintessence models.  In addition, the condition $c_s^2\ge0$ causes the 
barotropic dynamical behavior to be distinguishable from quintessence -- 
that is, they tend to lie in distinct regions of the equation of state 
phase space 
\cite{wSCH}.  Here we extend these results to a more general discussion of
the types of behavior that are allowed for barotropic models.  
We will see that limits $0\le c_s^2\le 1$ on the sound speed, along 
with some fairly general observational constraints, allow surprisingly broad 
conclusions to be made about the properties of viable barotropic models, 
ruling out some models in the literature.  Moreover, the properties describe 
an attractively simple picture of dark energy, together with a possible 
simultaneous resolution of the coincidence problem 
{\it and\/} why today the equation of state $w$ is near $-1$. 

We discuss general properties of barotropic fluids in \S\ref{sec:prop}, 
together with some special cases.  In \S\ref{sec:obs} we explore the 
distinction between barotropic and scalar field solutions to the dark 
energy puzzle, and the relation to \lcdm, finishing with a comparison 
of the coincidence and $w\approx-1$ behaviors of the different classes.

\section{Properties of a Barotropic Fluid \label{sec:prop}}

\subsection{General Properties \label{sec:gen}}

We define a barotropic fluid as any fluid in which the
physics of the fluid is fully determined by the 
pressure as an explicit function of the density:

\begin{equation}
\label{def}
p_{DE} = f(\rho_{DE}).
\end{equation}
Thus, the equation of state function $f$ completely characterizes
a barotropic fluid.  For example, the generalized Chaplygin gas
has the equation of state function \cite{Bento}
\begin{equation}
\label{Chap}
p_{DE} = -\frac{A}{\rho_{DE}^{\alpha}},
\end{equation}
where $A$ and $\alpha$ are constants, while the quadratic
equation of state is 
\begin{equation}
\label{quad}
p_{DE} = p_0 + \alpha \rho_{DE} + \beta \rho_{DE}^2,
\end{equation}
with the linear (or affine) model corresponding to the
special case $\beta = 0$.  The van der Waals equation of state
is \cite{VDW1,VDW2}
\begin{equation}
\label{VDW}
p_{DE} = \frac{\gamma \rho_{DE}}{1-\beta \rho_{DE}} - \alpha \rho_{DE}^2.
\end{equation}

How do barotropic models differ from quintessence models for dark energy?
For a canonical, minimally coupled scalar field $\phi$, the relation 
between pressure and density is given parametrically by
\begin{eqnarray}
p_\phi &=& \frac{1}{2} \dot\phi^2 - V(\phi), \\ 
\rho_\phi &=& \frac{1}{2} \dot\phi^2 + V(\phi),
\end{eqnarray}
where $\dot \phi = d\phi/dt$, and $V(\phi)$ is
the quintessence potential.

One can certainly find quintessence potentials for which
$p_\phi$ is not a single-valued function of $\rho_\phi$ and
which therefore can never be described in terms of equation~(\ref{def}).
For instance, a field oscillating about the minimum of a potential
has zero kinetic term at either extreme of the oscillation and 
zero potential term at the minimum (in terms of the equation of 
state ratio $w=p_\phi/\rho_\phi$, this is $w=-1$ and $+1$ 
respectively).  
Thus, $p_\phi$ passes through $0$ twice on every oscillation, with
a decreasing value of $\rho_\phi$ each time (since $\rho_\phi$ 
diminishes as the universe expands).  On the other hand,
many quintessence models can be characterized by a 
pressure which is a single-valued function of density, so what 
distinguishes these models from the barotropic models we
consider here?\footnote{Indeed, 
it is trivial to write down an effective potential for a barotropic 
model: 
\beqa 
V(\phi)&=&(\rho-f)/2 \nonumber \\ 
\phi&=&\int dt\,(\rho+f)^{1/2}\,, \nonumber 
\eeqa 
and from this $V(\phi)$ alone one could not tell if the physics was 
barotropic or quintessential.} 

The main distinction is that barotropic models have a value
of $dp_{DE}/d\rho_{DE}$ which is constrained by limits on the sound speed.
The sound speed for a barotropic fluid is given by
\begin{equation}
\label{sound}
c_s^2 = \frac{dp_{DE}}{d\rho_{DE}}.
\end{equation} 
Note that there is no need to write partial derivatives, since $p_{DE}$ 
depends only on $\rho_{DE}$.  To ensure stability, we must have 
$c_s^2 \ge 0$, so that
$dp_{DE}/d\rho_{DE} \ge 0$.  Thus, the function $f(\rho_{DE})$
in equation (\ref{def}) is not arbitrary; it must satisfy 
$df/d\rho_{DE} \ge 0$.  We will
further require, for causality, that $c_s^2 \le 1$ (\cite{Ellis}, but 
also see \cite{vikman}).  
This imposes the additional constraint $df/d\rho_{DE} \le 1$.

In contrast, a canonical minimally coupled scalar field is an imperfect
fluid.  While its adiabatic sound speed is given by equation
(\ref{sound}) (written as a partial derivative holding the entropy, 
or scale factor, fixed), its physical sound speed is always equal to the
speed of light.  Thus, these models are not subject to
the constraint that $dp/d\rho \ge0$; in fact, they generically
have $dp/d\rho < 0$ \cite{wSCH,CL}.  This is the reason that
barotropic models and many quintessence models occupy disjoint
regions in the $w-w^\prime$ phase plane \cite{wSCH}.
(For more general discussions of the behavior of scalar
field dark energy models in the
$w-w^\prime$
phase space, see \cite{CL,paths,Chiba,dynq}). 

Starting from the definition of the equation of state ratio, 
$w=p_{DE}/\rho_{DE}$, 
and taking the derivative with respect to the logarithmic scale factor 
$\ln a$, denoted by a prime, one has 
\beqa  
w'&=&-3(1+w)\left(\frac{dp_{DE}}{d\rho_{DE}}-w\right) \label{eq:wwp} \\ 
&=&-3(1+w)(c_s^2-w)\label{wwp2}\,. \label{eq:wpcs}  
\eeqa 
The requirement that $c_s^2 \ge 0$ then gives \cite{wSCH}
\begin{equation}
\label{wprime}
w^\prime \le 3w(1+w). 
\end{equation}
Here we consider only barotropic
models for which $w > -1$, although it is also possible
to generate barotropic phantom models with $w < -1$ 
\cite{stefancic04,odin0501,odin0508}. 
Since, for dark energy, $w<0$, we have $w^\prime < 0$ for
all barotropic fluids that can serve as dark energy.  Models
of this kind, in which $w$ approaches $-1$ with time, have been dubbed
``freezing" models \cite{CL}, although for quintessence freezing models 
one frequently has the opposite of equation~(\ref{wprime}): $w'\ge 3w(1+w)$.   
(Quintessence models can be found to violate this, while barotropic 
models will never break Eq.~\ref{wprime}).  
Further, the upper bound on the sound speed,
$c_s^2 \le 1$, gives a lower bound on $w^\prime$:
\begin{equation} 
\label{eq:null} 
w^\prime \ge -3(1+w)(1-w).
\end{equation} 
This is precisely the null line for the $w-w'$ phase plane \cite{paths} 
and leads to the one exception where quintessence models are exactly
equivalent to barotropic models: skating models \cite{sahlen,paths}, 
with kinetic energy along a flat potential, follow the equality in 
equation~(\ref{eq:null})  
and correspond to barotropic models with $f(\rho_{DE})=\rho_{DE}-\rho_\star$; both 
have $c_s^2=1$.  This is a pathological case, however, as the kinetic 
energy of skating models redshifts as $a^{-6}$ and the model quickly 
becomes indistinguishable from a cosmological constant. 

Since $w$ decreases with time, equation (\ref{eq:wwp}) has a generic future
attractor at $w=-1$, independent of the functional
form of $f(\rho_{DE})$.  At the attractor, the
density, pressure, and sound speed asymptotically
approach constant values, which we denote $\rho_\infty$,
$p_\infty$, and $c_{s\infty}$.  Taking $c_s = c_{s\infty}$ in 
equation~(\ref{wwp2}), we see that $w$ approaches $-1$ as
\beq 
1+w\sim a^{-3(1+c_{s\infty}^2)}, 
\eeq
and the dark energy density asymptotically approaches $\rho_\infty$ as
\begin{equation}
\label{rhoasymp}
\rho_{DE} - \rho_\infty \sim a^{-3(1+c_{s\infty}^2)}.
\end{equation}

Since at late times the density always approaches a constant, $\rho_\infty$,
this suggests decomposing the barotropic energy density into
two components,
\begin{equation}
\label{aetherdef}
\rho_{DE} = \rho_\infty + \rho_{AE}\,,
\end{equation} 
where the first term represents an always present cosmological constant 
and the second term defines an ``aether" density, $\rho_{AE}$, a 
spatially pervasive fluid with sound speed generally different from 
unity.  

The aether component is itself barotropic (with zero cosmological 
constant piece), and by comparing equations~(\ref{aetherdef}) and 
(\ref{rhoasymp}) to the usual behavior of some component $x$, 
$\rho_x\sim a^{-3(1+w_x)}$, one sees that $0\le w_{AE}\le 1$.  
That is, we can represent any barotropic fluid overall 
as consisting of a cosmological constant plus a positive equation 
of state perfect fluid.  The sound speed of the aether is equal to 
the sound speed of the full barotropic component, a result that is easily
proven using the formulas for summing components, 
equations (26) and (27) of \cite{dynq}, together with 
equation~(\ref{eq:wpcs}). 

Note that the cosmic acceleration from the 
dark energy arises purely from the cosmological constant piece. 
In contrast to the barotropic dark energy as a whole, the aether 
component acts to decelerate and has a density that decreases at least
as fast as the matter density, $\rho_M \sim a^{-3}$, and no
more rapidly than a stiff fluid, $\rho \sim a^{-6}$. 

As a specific example, consider the Chaplygin gas model, which is given by
equation~(\ref{Chap}) with $\alpha = 1$.
The full Chaplygin gas density evolves as
\begin{equation}
\rho_{DE} = \sqrt{\rho_\infty^2 + (\rho_{DE,0}^2-\rho_\infty^2)\,a^{-6}},
\end{equation}
with the future attractor at $\rho_\infty = \sqrt{A}$,
so the aether density
is given by
\begin{equation}
\rho_{AE} = \sqrt{\rho_\infty^2 + (\rho_{DE,0}^2-\rho_\infty^2)\,a^{-6}} - 
\rho_\infty\,.
\end{equation}
The aether density decays as $a^{-3}$ at early times (i.e.\ $w_{AE}=0$), 
while in the limit where the constant-density attractor is approached, 
$\rho_{AE}$ decays as $a^{-6}$ (i.e.\ $w_{AE}=+1$). 

Requiring a matter dominated era at high redshift constrains the 
behavior of the aether component at early times.  Since $w_{AE}\ge0$, 
the energy density of this component will tend to dominate in the past. 
However such an aether component with
a density greater than the matter density at high redshift
violates a host of observational constraints \cite{doran}.  
Thus, we must have $w_{AE} \rightarrow 0$ as $a \rightarrow 0$.
Furthermore, applying equation~(\ref{eq:wpcs}) to the aether 
component, we see that as $a\to0$ we have $c_s^2\le w_{AE}$ so 
in this limit $c_s^2\to0$. 

These arguments then allow us to put very general constraints
on the behavior of the overall barotropic dark energy:  it will always
behave like a pressureless dust component at early times, and like
a cosmological constant at late times.  (Note that ``late times"
could mean the far future, rather than the present).
In order to produce dust-like
behavior at early times, the functional form of $f(\rho_{DE})$ must satisfy
the constraint $df/d\rho_{DE} \rightarrow 0$ as $\rho_{DE} \rightarrow
\infty$, as well as the previously-discussed
sound-speed limits: $0\le df/d\rho_{DE}\le 1$ at all times.  We discuss 
the observational implications of this further in \S\ref{sec:obs}. 

For the
barotropic models we have mentioned, these limits impose severe constraints.
For example, for the quadratic and affine models of equation (\ref{quad}), our
limits impose $\beta = 0$ and $\alpha = 0$, so all acceptable
dark energy models of this
type reduce to the simple case $p_{DE} =$ constant.  This corresponds
to the special case where $\rho_{AE}$ behaves exactly like dust at
all times (see the next section). 
Similarly, the van der Waals model (equation~\ref{VDW}) is found to 
have unphysical behavior. 
For the generalized Chaplygin gas (equation \ref{Chap}), our constraints
give
$\alpha > 0$, ruling out several extensions of the 
generalized Chaplygin gas \cite{Sen}.

The aether decomposition also provides a simple recipe for producing
acceptable barotropic models.  Specifically, every barotropic equation
of state can be written in the form
\begin{equation}
p_{DE} = - \rho_\infty + g(\rho_{DE}-\rho_\infty), \label{eq:grho} 
\end{equation} 
where the function $g$ is subject to the constraints
\begin{equation} 
0 \le dg(\rho)/d\rho \le 1, \label{g1} 
\end{equation}
\begin{equation}
g(0) = 0, \label{g2} 
\end{equation}
and, in the limit where $\rho \rightarrow \infty$,
\begin{equation}
\label{g3}
dg(\rho)/d\rho \rightarrow 0.
\end{equation}

\subsection{Special Cases \label{sec:spec}}

Now consider some special cases of interest.
When $c_s^2 = 0$, we have the previously-mentioned
constant pressure model, characterized
by $p = -\rho_\infty$.  The density in this case evolves as
\begin{equation}
\rho = \rho_\infty + Ca^{-3},
\end{equation}
i.e., this looks just like the $\Lambda$CDM model (though with 
an additional matter contribution).  This
model has been previously discussed elsewhere.  For instance, it corresponds 
to the $\alpha = 0$ limit of the generalized Chaplygin gas \cite{Bento},
and it is a special
case of the ``mocker models" discussed in \cite{paths}, in
which it was noted that such models are characterized by
$w^\prime = 3w(1+w)$, defined there as the constant pressure line.  

For $c_s^2$ constant but nonzero, we have
\begin{equation}
\rho = \rho_\infty + Ca^{-3(1+c_s^2)} \,. 
\end{equation}
These models are all observationally excluded as noted earlier, since
they asymptotically dominate the expansion at high redshift.
In particular, pure skating models have $c_s^2=1$
and therefore correspond to $dg/d\rho = 1$ at all times; such models
then violate the condition in equation (\ref{g3}) above.

\subsection{Relation to k-essence \label{sec:kess}} 

Note that there is a one-to-one correspondence between
barotropic fluids and the subset of k-essence models with constant
potential, the so-called ``purely kinetic" k-essence models.  The
latter models were first investigated in the context of inflation \cite{Arm1}
and later proposed as unified models of dark matter and dark energy
\cite{Scherrerk}.  These models are
characterized by
a Lagrangian of the form:
\begin{equation}
\label{p}
p = F(X),
\end{equation}
where $X$ is
\begin{equation}
\label{grad}
X = \frac{1}{2} \nabla_\mu \phi \nabla^\mu \phi,
\end{equation}
and $\phi$ is the k-essence scalar field.
The pressure in these models is simply given by equation (\ref{p}), while the energy density is
\begin{equation}
\label{rho}
\rho = 2X (dF/dX) - F.
\end{equation} 

To convert between a kinetic k-essence model and a barotropic model, one 
inverts the function $F$ to find $X=F^{-1}(p)$ and substitutes this into 
equation~(\ref{rho}) to convert $\rho(X)$ into $\rho(p)$.  Inverting this 
delivers the barotropic equation~(\ref{def}).  If one starts with a barotropic 
model, one inverts $f$ to give $\rho=f^{-1}(p)$ and interprets 
equation~(\ref{rho}) as a differential equation to solve for $F(X)$.  To 
wit (cf.~\cite{kessence})  
\begin{equation}
\int\frac{dF}{f^{-1}(F) + F} = \ln(CX^{1/2})\,,
\end{equation}
where $C$ is a constant.  
This one-to-one correspondence (assuming the functions are invertible) 
means that the trajectories for purely
kinetic k-essence lie in the same region of the $w-w^\prime$ phase
plane as the trajectories for barotropic fluids \cite{dePL}. 

For the constant $c_s^2$ barotropic model, for example, the explicit 
barotropic relation between pressure and density is 
\beq 
p_{DE}=c_s^2\,\rho_{DE}+(1+c_s^2)\,\rho_\star, 
\eeq 
and the analogous k-essence Lagrangian is \cite{dePL,Quercellini} 
\beq 
p=\rho_\star+A\frac{2c_s^2}{1+c_s^2}\,X^{(1+c_s^2)/(2c_s^2)}, 
\eeq 
where $A$ is an arbitrary constant allowed by field redefinition in 
$X$. 

\section{Observational Signatures \label{sec:obs}}

Despite the generality of the models discussed here, they do provide
some distinctive observational signatures.  We first consider features 
in the homogeneous background properties, e.g.\ expansion history and 
equation of state, and then in the spatial perturbation properties. 

Consider first the jerk parameter of the expansion history \cite{visser} 
\begin{equation}
j \equiv \frac{a^2\dddot a}{\dot a^3}.
\end{equation}
The importance of this parameter was first emphasized by 
\cite {CN}, and it was presented as one 
of the ``statefinder'' parameters in \cite{state1,state2}.  
In a flat universe, $j$ is given
by \cite{CN}
\begin{equation}
j = 1 + \frac{9}{2}\Omega_{DE}\frac{dp_{DE}}{d\rho_{DE}}(1+w_{DE})\,. 
\end{equation}
Thus, for $\Lambda$CDM (including the constant pressure models 
discussed in \S\ref{sec:spec}), we see that $j=1$, independent
of the value of $\Omega_\Lambda$ \cite{Blandford}.  For
the barotropic models considered here, the requirement that
$dp_{DE}/d\rho_{DE} \ge 0$ translates into a simple bound:  $j \ge 1$.
On the other hand, for quintessence models, we can write \cite{wSCH,paths}
\begin{equation}
j = 1 - \frac{3}{2} \Omega_{DE}[w^\prime - 3w(1+w)].
\end{equation}
Since quintessence models generally satisfy
$w^\prime \ge 3w(1+w)$ \cite{CL}, we see that they are often characterized
by $j \le 1$.  Thus, accurate measurement of jerk parameter $j<1$ 
(or equivalently $w$ and $w'$) could provide perhaps the cleanest 
observational signature to 
distinguish barotropic dark energy from quintessence. 
One can view this as determining the number of internal degrees of 
freedom in the dark energy physics \cite{Quercellini,caldpriv}. 

Unfortunately, the jerk parameter is rather difficult to derive 
from current observational data.  Assuming a constant value for $j$, 
\cite{Rapetti} 
derived $j = 2.16^{+0.81}_{-0.75}$.  While this might na{\"\i}vely
seem to favor barotropic models, in fact the assumption of constant
$j$ strongly biases the result.  Typical barotropic and quintessence
models generally have values of $j$ that vary significantly with time,
and current observations are insufficient to distinguish these two types
of models.

Note that as $j$ approaches unity, it becomes difficult to distinguish 
barotropic (or quintessence) models from \lcdm\ using the expansion 
history, since the constant-pressure model is degenerate with $\Lambda$CDM.  
Two possible observational signatures exist.  In the case of barotropic
dark energy, some of the contribution to the zero-pressure dark component can
arise not only from
dark matter but also from the aether component.  
If the dark matter particle is detected, and its relic abundance
can be calculated from its physical properties, then
one signature of barotropic dark energy would be an anomalously
high observed value of $\Omega_M$ in relation to the dark matter 
(and baryon) calculation.  

A second possible signature is the behavior of spatial perturbations.
Since $c_s^2=0$ for the constant-pressure barotropic model, one might 
hope to distinguish it 
observationally from the corresponding quintessence model 
with $c_s^2 = 1$ on the basis of perturbation growth.  (Constraints on
the dark energy sound speed have been explored for constant sound 
speed in \cite{Bean,Hannestad}). 
Interestingly, where barotropic and quintessence models are closest 
in their dynamics (i.e.\ near $w'\approx 3w(1+w)$), they differ most in 
sound speed, and where they are closest in sound speed, they differ 
most in their dynamics.  
However, as $w\to-1$, barotropic models and \lcdm\ become degenerate, at least 
to linear order.   We can see this by 
examining the equation for linear perturbation growth in the
dark energy fluid in synchronous gauge \cite{VS,Avelino}:
\begin{equation}
\label{delta}
\dot \delta_{DE} + (1+w)(\theta_{DE} + \dot h/2) + 3H(c_s^2-w)\delta_{DE} = 0,
\end{equation}
where $h$ is the trace of the perturbation to the Friedman-Robertson-Walker
metric, and $\theta_{DE}$ is the divergence of the fluid velocity.  The key 
point is that for the case $w=-1$, there is no growing mode in 
$\delta_{DE}$, since the last term is always positive.  
Thus, $w=-1$ models cannot be distinguished, regardless of the value of
$c_s^2$ (this point is emphasized in \cite{Avelino} for the
case of the $\alpha=0$ Chaplygin gas and $\Lambda$CDM).  

One can also argue qualitatively from equation (\ref{delta}) that 
the growth of density perturbations should become relatively
insensitive to $c_s^2$ in the limit where $w$ is close to $-1$.
This conclusion is borne out by detailed comparisons in the $w - c_s^2$ plane 
between models and the observations \cite{Bean,Hannestad}. 
These investigations show likelihood curves that are nearly 
independent of $c_s^2$ for $w$ near $-1$. 

Note that barotropic models as a class can never be ``ruled out" 
as long as $\Lambda$CDM remains viable, since as a limiting case 
constant-pressure barotropic models include $\Lambda$CDM.  
We have shown ways, however, in which one can generally constrain 
the allowed parameter space for barotropic models.  That said, 
all observationally-allowed 
barotropic models must approach the constant-pressure model at 
high redshift, making them indistinguishable from $\Lambda$CDM 
in this limit (save through the theoretical predictions
for $\Omega_M$ noted above). 

One can take this further and 
argue that barotropic models ``predict" a value of $w$ near
$-1$, in a way that quintessence models do not.  Our argument is based
on the upper bound on $w'$ given by equation (\ref{wprime}).  This
equation shows that, for barotropic models with nonnegative $c_s^2$,
the value of $w$ for the dark energy can never ``loiter" at a
value between $w=0$ and $w=-1$.  While $w$ can lie near $0$ for
arbitrarily long times in these models, once it begins to decrease
toward $-1$, equation (\ref{wprime}) puts a lower bound on the rate
of decrease.  Thus, one cannot have arbitrarily long periods in which
$w$ has some value between $0$ and $-1$.  The opposite is true
in quintessence models; it is easy to construct such models (trackers) with
$w$ roughly constant and equal to nearly any desired value \cite{Stein1}.

In terms of our aether decomposition, the slowest rate of decrease
for $w_{DE}$ occurs when $w_{AE} = 0$.  In this case, for example,
$w$ decreases from $-0.1$ to $-0.9$ as the scale factor increases
by about a factor of 4 (i.e.\ within 1.5 e-folds).  
Other choices for the equation of state
function can only produce a more rapid decrease in $w$.  Thus, in barotropic
models, a value of $w$ between $0$ and $-1$ must always be a transient
phenomenon, leading to the argument that barotropic models ``predict"
a value of $w$ near $-1$ (a prediction which would have been considerably
more convincing had we made it a decade ago).  Note that a $w \rightarrow -1$
attractor is also present in some unified models for dark matter and dark
energy \cite{U1,U2}, although such unified models are outside the scope
of our discussion.  

Since the only viable barotropic models are those that scale like the 
dominant component at high redshift, no other special selection needs to 
be applied: in this sense the ``bug'' of not being able to distinguish 
a constant-pressure barotropic model ($w_{AE}=0$) from \lcdm\ is 
really a ``feature'' of ameliorating the problem of fine tuning initial 
conditions.  (The 
usual cosmological constant problem remains of why the non-aether part, 
$\rho_\infty$, is so small.) 

Note that the coincidence problem is also somewhat
ameliorated (see \cite{eganline} 
for a recent discussion of the coincidence problem from a novel 
perspective).  For the case of a cosmological constant, the
ratio of dark energy density to matter density has to increase 
by nine orders of magnitude between 
recombination and today, leading to the question of why dark energy
overtakes matter 
basically now.  In contrast, the barotropic models can easily have a more 
natural-seeming ratio of order $10^{-2}$ or ${\mathcal O}(1)$ at 
recombination.  While 
quintessential tracking models can also have such ratios, they have 
difficulties in then achieving $w\approx-1$ today.  Thus barotropic 
models have attractive characteristics with regards to both the fine 
tuning and coincidence problems.  In effect, the aether 
component of the barotropic fluid anaesthetizes the cosmological constant 
against the pain of fine tuning.

\section{Conclusions} 

Barotropic fluids have a number of characteristics that make them an 
interesting class of dark energy models.  They have, by definition, 
an explicit rather than implicit equation of state relating the 
pressure and the energy density.  While this relation is nominally 
quite general, we show that simple physical conditions such as stability 
and causality severely restrict the allowed functional forms.  In 
particular, we demonstrate that viable barotropic models must 
possess the following properties: 
\begin{itemize} 
\item Asymptotic future de Sitter state, where the dynamics freezes to a 
cosmological constant state, 
\item Dynamics distinct from much of quintessence, lying in a separate 
region of $w$-$w'$ phase space; sound speed generally distinct from 
quintesssence, 
\item Acts as a sum of a cosmological constant and a perfect fluid 
``aether'' component with $w_{AE}\ge0$, 
\item The aether component must have $w_{AE}\to0$, $c_s^2\to0$ in the past 
in order not to violate matter domination. 
\end{itemize}
These results both unify a number of special cases in the literature 
and rule out several models.  

We consider several observational signatures to distinguish barotropic 
fluids through both the effects on background expansion and on perturbation 
growth.  For example, in the barotropic case the sound speed is not 
restricted to be the speed of light, as in canonical, minimally coupled 
scalar field models.  Constant-pressure ($c_s^2=0$) barotropic models are 
however degenerate with \lcdm, though possibly distinguishable through 
a discrepancy between particle physics predictions for the dark matter 
density and cosmological observations. 

Finally, the aether component of the barotropic fluid can anaesthetize 
the cosmological constant against some of its fine tuning and coincidence 
problems.  In the high redshift universe the dark energy appears like 
\lcdm, but with a dark component 
energy density that can be comparable to the 
matter density.   At late times, it naturally and rapidly transitions
from a matter-like behavior to 
behavior that approaches a pure cosmological constant.

\acknowledgments 

E.V.L. was supported in part by the Director, Office of Science, 
Office of High Energy Physics, of the U.S.\ Department of Energy under 
Contract No.\ DE-AC02-05CH11231.
R.J.S. was supported in part by the Department of Energy (DE-FG05-85ER40226).

{}

\end{document}